\documentclass[12pt]{article}          

\usepackage{cite}
\usepackage{amsmath,amssymb,amsfonts}
\usepackage{algorithmic}
\usepackage{graphicx}
\usepackage{textcomp}
\usepackage{xcolor}
\usepackage{multirow}
\usepackage{caption}
\usepackage{subcaption}
\usepackage{fullpage}
\def\BibTeX{{\rm B\kern-.05em{\sc i\kern-.025em b}\kern-.08em
    T\kern-.1667em\lower.7ex\hbox{E}\kern-.125emX}}
\begin{document}

\title{Redefining measures of Layered Architecture}

 \author{Sanjay Thakare$^1$, Arvind W Kiwelekar$^2$\\
Department of Computer Engineering \\
Dr. Babasaheb Ambedkar Technological University,\\Maharashtra, India.\\
1:mail2sbt@gmail.com,2:awk@dbatu.ac.in}

\maketitle

\begin{abstract}
	Layered architecture represents the software structure in the form of layers. Every element in the software is assigned to one of the layers such that the relationship amongst the elements is maintained. The process of construction of the layered architecture is ruled by a set of design principles. Various statistical measures have been defined to check whether the layered architecture of a given software is following these design principles or not. In this paper, we redefine the measures of layered architecture based on the relationship between the software components. The measures check for the violations committed in terms of the back calls, skip calls, and cyclic structures. Further, we also introduce a new measure to verify the logical separation amongst the layers. The current architecture of the system is extracted from the source code and represented using a three-tier layered structure, which is the defacto standard architecture of Java applications. The redefined measures are applied to determine the conformance of layering principles in the system. We evaluate five different software systems for their architecture consistency. The results obtained on our redefined measures are compared to those obtained by applying the standard set of measures. A quantitative analysis of the proposed measures is performed and we conclude that they have the potential to determine the consideration of layering principles followed during the development of a software system.
	 
\end{abstract}


\section{Introduction}
\par Source code analysis techniques and evaluation of architecture have been studied for decades, providing multiple views of the software and revealing design decisions that were considered during the development of a project. They provide a significant amount of information to improve the performance of the legacy software systems and to make decisions regarding the refactoring and maintenance. The software is often monitored to understand the impact of ad-hoc changes due to the fixation of a bug, or the addition of new features to it. A perspective is also provided to the system in terms of maintenance, performance, and evolution.

\par Layered architecture is a simple, common yet powerful architectural style used in most of the application design. Effective recovery and evaluation of the layered architecture is a problem that still poses a profound challenge to the researchers. This is due to the liberty present with the developers to take decision particularly suitable for certain situations, and the ad-hoc changes made in the software due to peer pressure. For designing of layered architecture, the system is often broken down into a set of layers such that each layer has a well-defined role and responsibility. Now, the communication pattern imposed by layering structure promotes a gradual reduction of computing complexity, so that a higher layer consumes the services provided by the lower ones to create a complex functionality. The individual components present in the layers are interdependent, thus forming a network structure. A layered architecture can be evaluated based on certain measures that check whether any predefined design constraints have been violated or not.

\par The hierarchical structure is a result of the intentionally created hierarchical dependency relations among the components. These are defined and promoted by the layered structure. The structure is also represented and studied in the form of the dependencies network or graph. As the complexity and size (the number of components) of the software increase over the period, the network size also increases and the network becomes complex for analysis. Therefore,  we required a sophisticated technique to handle such a large and complex network. Social Network Analysis (SNA) is a widely used network analysis technique that primarily focuses on the relationships that exist in large and complex networks.

\par Social network is fundamentally very similar to a software dependency network, in which a set of software components are considered as social entities, and dependency among the software component is a social relation. Researchers have applied the concepts and methods of social network analysis to understand the software systems and have observed that the network structure of the software system has resembling properties. In this study, we consider a software dependency structure as a social network. Therefore, measures that evaluate the layered structure need to be redefined in terms of this new relation perspective.

\par In this paper, we have redefined the measures of layered architecture evaluation defined by Sarkar \cite{sarkar2009discovery} in terms of relation to evaluate the layered structure. Adding to this, we have proposed new metrics that measure the logical separation of the layer. We measure the extent to which the layers are logically separated from each other to evaluate the recovered layered architecture. In this study we address the following issues:
\begin{enumerate}
    \item Monitoring the status of the software.
    \item Keeping a check on the design considerations of the software.
    \item Check the existence of the layered structure, and analyzing its nature (open or closed).
\end{enumerate}

\par The paper structure is as follows: Section II presents a brief overview of the related work in this domain. The redefined measures in terms of dependency relationships are explained in section III. Experimental results obtained on a sample network and two real-life software systems are presented in section IV and we conclude our findings in section V.

\section{Related work}
\par In the past, researchers have emphasized and provided significant attention towards understanding the evolving software system. They have relied on their domain knowledge and some predefined algorithms to update missing information that is not present in the original specification. Design specifications can be ignored in the case of small scale systems to again restart from scratch in case of evolution of software. However in case of large-scale and complex systems, the inputs from previous versions are very much needed to produce a more relevant and up-to-date architecture. Such high level recovered architecture designs are a good way to understand a legacy software and the involved dependencies before moving on to making new changes \cite{murphy2001software}.
\par Reverse engineering in case of software systems helps to reduce effort of developers for modernizing, and understanding the complex systems where recovery of high-level structure from the source code is expected. It is necessary to evaluate the recovered structure to measure the drift from the specified design documents or design patterns. Our work focuses on the evaluation of recovered layered architecture from the object-oriented software system. 
There are various method for layered architecture recovery proposed by researcher based on various heuristics \cite{belle2013layered},\cite{belle2014recovering}, \cite{laval2013ozone}, \cite{sarkar2009discovery}, \cite{scanniello2010architectural}.
Various metrics have been proposed to evaluate the recovered layered architecture. But our work is based on the measures proposed by \cite{sarkar2009discovery}. 
\cite{lague1998analysis} studied interfaces (.h header files) in layered architecture of telecommunications systems. Further, they proposed a set of metrics and analysis dependency graphs to answer a set of specific questions for the evaluation of the properties of a layered architecture. The set of metrics used to measure the interface were as follows:
\begin{itemize}
    \item \textit{Up Layer Interface}: The ratio of files from all the top layers referred to the files of a specific layer.
    \item \textit{Down Layer Interface}: The ratio of files in a specific layer to the files present in all the bottom layers. 
    \item \textit{Layer/Layer Up Layer Interface Ratios}: The ratio of files from the top layer referred to the files of a specific layer.
    \item \textit{Layer/Layer Down Layer Interface Ratios}: The ratio of files in a specific layer with respect to the files from a particular down layer.
    \item \textit{Services Related Files Ratios}: The ratio of files in a specified layer that provides services to another layer, to the total files in that layer.
\end{itemize}

\par However, these interface metrics are unable to provide a complete evaluation of the layered architecture as they look upto every layer in an independent manner with no comments on any inter or intra-dependencies. Similarly, Sarkar \cite{sarkar2009discovery} defined different metrics to evaluate the layered architecture recovered from source code. The layered architecture was created using a clustering algorithm and improved with the help of a domain expert. They designed various index measures to check the extent to which a layered structure violates the design philosophy. 

\par Ran Mo \cite{mo2016decoupling}  proposed a metric to measure architecture maintainability DL-\textit{decoupling level} which focused on the decoupling of program elements instead of measuring coupling level. The study claimed that modularity creates value in the form of options and that independent and replaceable modules increase the system value. The independent modules were identified using the Design Rule Hierarchy (DRH) clustering algorithm and were then used to calculate the decoupling level of a system. DL value remains the same for a non-refactoring and unchanged snapshot of the project and indicates either improvement or degradation of the relative changes in the system DL. It also indicates the extent to which the architecture is logically separated into small, independent modules that can be revised or developed in parallel by developing a team.

\par Complex network theory or Social Network Analysis provides sophisticated concepts, methods, techniques and tool to analyze various kinds of networks\cite{knoke2019social}. It has been successfully applied for analysis of structure of different real-life systems across various domains such as social science, physics, biology and computer science. Gradually, this exploration also extended to computer software systems. Even though software systems vary in terms of development, they exhibit similar properties to those of other real-world systems \cite{de2003signatures}, \cite{hyland2006scale}, \cite{valverde2002scale}, \cite{myers2003software}.
In case of software network, the nodes could be files, methods, classes and packages where as edges could indicate the associations or dependencies. Such a work is not limited to verification of the properties of real-life systems, but also involves extensive study of property like community \cite{vsubelj2011community} for software system. Suggestions have also been made towards application of complex network theories in the field of software engineering \cite{vsubelj2011community}.  

Bhattacharya \cite{bhattacharya2012graph} used measures of complex network theory and link prediction technique to predict defect in evolving software system. In same line, Nguyen \cite{nguyen2010studying} verified previous work that claim that  measure of complex network theory are better in predicting the failures after release when combined with complexity metrics. Many research explored the application of these measures to predict the software defects. At same time, Schreiber \cite{schreiber2020social} made a comprehensive review on the latest SNA applications in software development projects.
This shows that researchers are taking interest in applying the complex network theory to software systems for enhancing the understanding of software systems.

\par This has been the motivation for us to study, apply, and analyze the techniques from complex network theory to software dependency networks for the recovery of layered architecture. The network theory is fundamentally based on the concept of relation amongst pair of nodes. Hence there is a necessity of a measure that is defined in the same terms. We can analyze the software network, recover and study the layered architecture from software dependency structure using Social Network Analysis.

\par Our effort in this paper is to define measures for architecture evaluation that align with network theory. Therefore, the aim of our work is to redefine the measures of layered architecture, so that we can use these measures to validate results produced by future works, involving recovery of layered structure by the application of network theory to various software systems. The main reason for redefining the current measures are that they only identify the modules which violates the principles, whereas in our work, we also identify those relationships that violate the principles. 

\section{Measures of layered architecture in terms of relation}
\par We have redefined measures for detecting design flaws in terms of the relation amongst nodes. A relation is an association of some sort like dependency, or interaction between two entities. Naturally, it is a mechanism of connecting them, and design is a way of organizing these connections to achieve certain goals. When things are interconnected, we can study them to consider either an isolated individual or the relationship. Social Network Analysis gives more attention to such kind of bonds or agreements between the individual nodes.
\par In this study, we primarily focus on the relationship that is the connection between the software entities and make a sincere attempt to redefine the measures for the evaluation of layered structure. Sarkar et al. \cite{sarkar2009discovery} illustrated measures in terms of software entity i.e. module. These measures capture the three fundamental dimensions of violations: back-call, skip-call, and cyclic violations. It describes a violation in term of a module which initiated the call and defines an index for it.
\begin{itemize}
    \item The back call principle states that the upper layer should only depend on the immediate lower layer and there should not be any reverse dependency. A top-down approach is thus expected to be followed during the design process.
    \item The skip call principle states that the dependency should exist from a top layer to any down layer. This helps to avoid duplication of functionality.
    \item The cyclic dependency is a special case of a back call that allows the formation of cycles of dependencies amongst the layers. It leads to a vast increase in the complexity of the layers while reducing their flexibility. It is preferred to be non-existent in the architecture.
\end{itemize}

\par The major limitations with these measures and index are as follows:
\begin{itemize}
    \item The index provides generic information of violation as it only considers the ratio of modules violating a particular rule to the total modules under consideration.
    \item It is unable to describe the severity of violations. No predictions or comments are made about the severity of the violations committed.
    \item It only considers the module which makes a call. However, this may not always true and the violation may have been committed by either of the caller, called, or both the modules in the sense that both have certain stakes in it.
    \item In the case of cyclic violation, \cite{sarkar2009discovery} considers all the arches that crosses a layer boundaries are responsible for the violations. However, in some cases these arches which crosses the layer boundary, are significantly necessary, and needed to follow the communication pattern imposed by the architecture.
\end{itemize}

\par In our approach, we are also able to detect the layering principle violation at various levels of granularity. Generally, the violation focuses on a relationship that crosses the layering boundaries. Layering allows a layer to interact with the immediate lower layers, and only such crossings are required and permitted. We now introduce a set of notations used throughout this paper:
\begin{enumerate}
	\item Let $G=\{P,E\}$ be the directed graph or network of software entities.
	\item Let $P= \{p_1,p_2,....,p_m\}$ be the set of all the program elements(modules or packages of a Java-software system) and $m = |P|$ be  total number of program elements in the system.
	\item Let $L=  \{l_1,l_2,....l_n \}$ be the set of layers recovered or in which the system is divided and $l=|L|$ be total number of layers such that $l>0$.
	The topmost layer is $l_i$ and $i=1$, whose numbering starts from 1 and increases gradually as we go downward.
	For any two immediately adjacent layer $l_i <l_j$, indices are such that $0<i<j<n$ and $j=i+1$. 
	\item  The program element(module or package) assignment function $f(P):l$ or $P->l$,  defines a mapping of a program element in P to a layer $l$ in $L$. This is a many-to-one function.
\end{enumerate}
\par Besides these, some useful notations are defined and used throughout the paper. Two non-trivial functions give a source and destination node of the arch 'e'.
\begin{equation}
\centering
src = \textit{S}(e)     
\end{equation}
\begin{equation}
\centering
dest = \textit{D}(e)
\end{equation}

\par The function determines back-call violation of arch 'e' iff $ \ell(src) > \ell(dest)$. Further, an arch 'e' violates skip-call iff $|\ell(dest) - \ell(src)| \geqslant 2 \quad  and \quad \ell(src) < \ell(dest)  $ The function $\mathcal{L}(m)$ determines the layer of a program element p:
\begin{equation}
\mathcal{L}(p) = l.    
\end{equation}

\begin{figure}[h]
	\centering
	\includegraphics[width=0.7\linewidth, height=0.3\textheight]{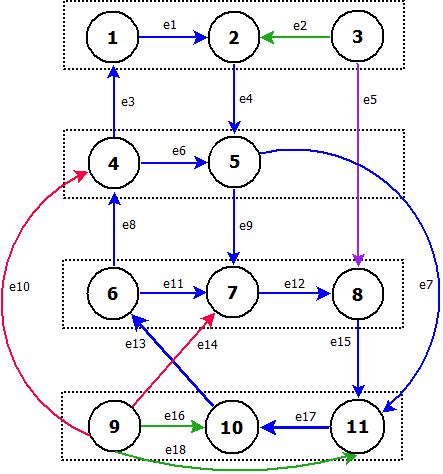}
	\caption{Sample network of the program elements with cycle.}
	\label{fig:basediagram}
\end{figure}

\subsection{Measuring Back-call Violation}
Let BV be the set of arches (i.e. edges) that violate the back-call principle.
It is expressed as follow:
\begin{equation}
BV = \{ \exists e \in E| \ell(src) > \ell(dest)\}
\end{equation}
\par $BV=\{e_{3}, e_{8}, e_{10}, e_{13}, e_{14} \}$ as shown in  Figure \ref{fig:basediagram}. If we apply the function $S$ as defined above, we get a set of caller modules which is equal to $BACK$ defined in \cite{sarkar2009discovery}, so 
$BACK = S(BV)=\{4,6,9,10\}$. Further, the called modules are obtained by using the function $D$ and $D(BV)=\{1,4,6,7\}$.

\par Back-call violation for a caller program element $m$ is a ratio of the arches that originate from $m_s$ in the set $BV$ to all arches originated from $m_s$. Similarly, for the called program element, we can define $m$ as the ratio of arches arriving at $m_d$ in the set $BV$ to all arches arriving at $m_d$.
\begin{equation}
BVM_{caller}(m_s)=\frac{|\{\exists e \in BV| \textit{S}(e) = m_s \}|}
{|\{\exists e \in E| \textit{S}(e) = m_s\}|}     
\end{equation}
\begin{equation}
BVM_{called}(m_d)= \frac{|\{\exists e \in BV| \textit{D}(e) = m_d \}|}
{|\{\exists e \in E| \textit{D}(e) = m_d\}|}    
\end{equation}
\par Further, we can write a back-call violation for an entity as an average of the above equations as follows:\\
\begin{equation}
BVM(m)= \frac{|\{\exists e \in BV| \textit{S}(e) or \textit{D}(e) = m \}|}
{|\{\exists e \in E| \textit{S}(e) or \textit{D}(e) = m\}|}
\end{equation}
\par This equation tries to capture the severity of a program element in the context of violations present in the layer. It is a fraction of arches (incoming and outgoing) that violate the rule to the total arches incident on the program element. One can observe that the denominator is the sum of in-degree and out-degree of the program element. The above expression can be further simplified as follows:
\begin{equation}
bcc = \{\exists e \in BV| \textit{S}(e) = m \}
\end{equation}
\begin{equation}
bcr = \{\exists e \in BV| \textit{D}(e) = m \}
\end{equation}
\begin{equation}
BVM(m) = \frac{|bcc|+|bcr|}{Deg(m)}
\end{equation}

\par Here, the terms $bcc$ and $bcr$ are the numerator of above expression.

\par Next, we define the back-call violation for the whole layer which is an average of back-call violation of all program elements.
In other words, it is a ratio of the arches that violates the back-call to all incident arches on the layer. 
\begin{equation}
BVL(l)= \frac{|\{\exists e \in BV| \mathcal{L}(src)  or  \mathcal{L}(dest) = l \}|}
{|\{\exists e \in E| \mathcal{L}(src) or \mathcal{L}(dest) = l\}|}
\end{equation}
\par The quantity at the denominator, in this case, could be determined by using ${Indegree}+ {Outdgree}+ 2 * \; \text{arches within a layer} $. Here, one should notice that in the context of a layer, there are fundamentally three types of arches based on its orientation. Some of them are incoming to the layer, others leaving from the layer and the rest fall within the layer.
\par Now back-call a violation of the entire system is defined by taking a ratio of $|BV|$ and $|E|$.
\begin{equation}
BVS(S) = \frac{|BV|}{|E|}
\end{equation}

\par Here, the numerator and denominator are multiplied by two due to the incident relation of an arch, hence each arch is counted twice, first in the form of outgoing arch and second as incoming.
\subsection{Measuring skip-call violation} 
Let $SV$ be the set of arches that violate the skip-call principle expressed as follow: 
\begin{equation}
SV = \{ \exists e \in E| |\ell(dest) - \ell(src)| \geqslant 2\}
\end{equation}	
provided that $\ell(src) < \ell(dest)$.
For the sample example shown in Figure \ref{fig:basediagram}, $SV=\{e_{5},e_{7}\}$.	
Skip-call violation for a program element m :
where 	$scc=\{\exists e \in SV| \textit{S}(e) = m_s \}$ be a set of arches leaving from m, and  a set $scr=\{\exists e \in SV| \textit{D}(e) = m_d \}$ represent incoming arches to m.
Consider the degree of the m as follows:
$InDeg(m)=|\{\exists e \in E |\textit{D}(e) = m_d  \}|$, $OutDeg(m)=|\{\exists e \in E |\textit{S}(e) = m_s \}|$ and $Deg(m)=|\{\exists e \in E |\textit{S}(e) = m_s \; or \; \textit{D}(e) = m_d  \}|$.
\begin{equation}
SVM_{caller}(m_s)= \frac{|scc|}{OutDeg(m_s)}    
\end{equation}
\begin{equation}
SVM_{caller}(m_d)= \frac{|scr|}{InDeg(m_d)}
\end{equation}
So, overall skip-call violation for a program element $m$:
\begin{equation}
SVM(m) = \frac{|scc|+|scr|}{Deg(m)}
\end{equation}
An average of skip violation of all program elements present in a layer is indicated by $SVL(l)$. In other words, it can be defined as the ratio of arches that violates the skip-call to all the incident arches on the layer. So, skip-call violation for a layer is as follows:	
\begin{equation}
SVL(l)= \frac{|\{\exists e \in SV| \mathcal{L}(src)  or  \mathcal{L}(dest) = l \}|}
{|\{\exists e \in E| \mathcal{L}(src)  or \mathcal{L}(dest) = l\}|}
\end{equation}
The denominator is determine by using ${Indegree}+ {Outdgree}+ 2 * \text{arches within a layer} $.\\
Skip-call violation of the entire system is derived as follows:\\
\begin{equation}
SVS(S) = \frac{|SV|}{|E|}
\end{equation}
\subsection{Detecting cyclic violation}
\par Let $C= \{c_{1},c_{2},......,c_{k}\}$ be the set of cycles present in a network. We assume that a network is a connected component, if not, then there exists a connected component in the network such that nodes are reachable from each other. Sarkar et al. \cite{sarkar2009discovery} defined an index to measure the cyclic violation, for a given a strongly connected component, as a ratio of the number of edges that across layers to the total number of edges that exist in the strongly connected component.
\par A cycle happens because of the completeness of the path, which means it ends to where it started. However, if a cycle is present in the layered architecture, then categories into two types:  within a layer or across a layer. In case of cycle cross the layer then the arches are 1) within the layer, and 2) between the layers. One of the non-trivial inter-layer from the context of cyclic dependency is permissible by the structure that involves arches in forwarding direction (except skip-calls). But arches making backward calls are a principle contributor to the cyclic violation.
\par Hence, we define a measure that is a fraction of the arches in the cycle that violate the rules imposed by the layered structure. We consider the maximal strongest connected component for measuring cyclic violation of a system. Here $CBS$ represents arches in a cycle that violate design rules of the layered structure.

\begin{equation}
CBS = \{ \exists e \in  E_{c_i}| \text{e violates back-call or skip-call}\}    
\end{equation}

\par Here, $c_i$ is the maximal strongest connected component of the network. Therefore, cyclic violation of entire system determined from maximal strongest connected component is as follows:
\begin{equation}
CV(S) = \frac{|CBS|}{|E_{c_i}|}
\end{equation}
Where $E_{c_i}$ is a subset of arches present in maximal strongest connected component.
\subsection{Average system violation}
The violation of entire system including both back-call and skip-call is defined as:
\begin{equation}
ASV = \frac{\alpha|BV|+ \beta |SV|}{|E|}
\end{equation}
Here, $\alpha$ and $\beta$ are penalties in the range $[0..1]$ for each arch that violates the architecture rules imposed by layered architecture.
But we know that some of the arches from BV also contribute to the cyclic violation of the system, hence, it will make the system tightly coupled and hard to maintain. 
Therefore, for such arches, we charge double penalty and above equation is defined as:
\begin{equation}
ASV =  \frac{(2\alpha |CB|+ \alpha|RB|)+ (2\beta|CS|+\beta|RS|)}{|E|+|CB|+|CS|}
\end{equation}	
Where $CB$ and $CS$ are the set of arches that present in the cycle and $RB$ and $RS$ are remaining arches from the set $BV$  and $SV$ after subtracting $CB$ and $CS$.
\subsection{Logical separation index}
Each layer encapsulates a logical, high-level responsibility of the system. It consists of a large number of program elements(modules, packages, sub-subsystem, etc.) that are interdependent. A system as a whole provides a set of functionalities where these dependencies pay a non-trivial role. Each layer, perform its responsibility by using services of various layers and also provides service to others. If a system follows layered architecture then a layer uses service from layers beneath it and provides services to the top layer. So, this means there is a clear separation of responsibility of a layer. 
The logical separation of a layer is measured as:  
\begin{equation}
LS(l)=\frac{\text{No of arches violates back-call principle} }{InDeg(l)+OutDeg(l)}
\end{equation}
\begin{equation}
LSI(l)=1-LS(l)
\end{equation}
Here, index value is in between $[0..1]$ where the index value 1 indicates that the layer is logically well separate and 0 means logically non-separable.
\section{Experiment Results}

\begin{figure}
	\centering
	\includegraphics[width=1\linewidth, height=0.3\textheight]{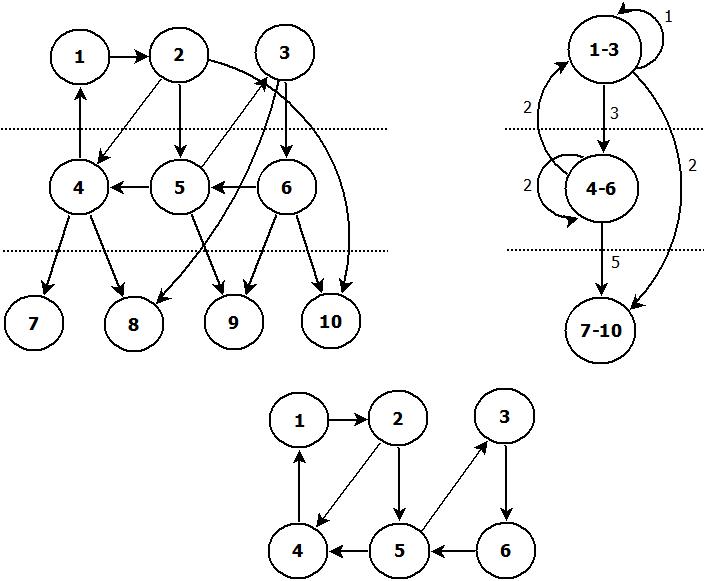}
	\caption{a) Sample network 2 organized into layered structure, b) layered dependency, c) maximal strongest connected component.}
	\label{fig:sampleexample}
\end{figure}

\par We consider a simple example as shown in Figure  \ref{fig:sampleexample}, in which the first figure represents the sample network, and the second figure represents a network of the layered structure of the example. The sample network consists of 10 nodes and 15 arches distributed into three layers. In this example, we have network configuration in a layered manner, nodes 1, 2, and 3 are placed in the top layer, node 4, 5, and 6 in middle layer while remaining nodes like 7, 8, 9, and 10 in the lower or bottom layers.

\par In this section, we describe the result of an experiment, where we applied our redefined measure to analyze the layered architecture of various software systems including sample network. First, we applied these to the sample network from Figure \ref{fig:sampleexample}a, Table \ref{tab:archanalysis} shows the analysis of the arches in the network. We found that two arches from the middle layer refer to the top layer and two from the top layer access services from the bottom layer. This indicates that four arches from the network don't follow the rules imposed by a layered architecture while remaining arches are either fall within the layer or between the layers that forward call.

\par The detailed analysis of a sample network is carried out at node (module or package) level. Any arch incident on the node is either normal, back or skip-call. A normal arch means it is within a layer or follows the rules of layered architecture. However, back and skip arch violates the rules of layered architecture. We have analyzed the node from two different perspectives:
\begin{enumerate}
    \item It makes a call to other node refereed as caller node.
    \item It get calls from other node referred as called node.
\end{enumerate}

\par We agree with \cite{sarkar2006method} that caller node is important in the violation analysis, in the sense that it establishes a connection to the interface and is a consumer of service offered through interfaces. However, we have also taken into account called nodes for analysis as it is indeed a non-trivial partner of the relation which may follow or violate the rules. Fundamentally,  somehow a called node is related and contributes to the relation that must not be ignored.  
\par A node violates rules means either it has leaving arches or incoming arches that don't follow the prescribed rules as shown in Table \ref{tab:back-callandskip-call}. The middle layer makes two backward-call to the upper layer, trying to access services from the above layer result in tightly coupled and fragile architectures that are very difficult to test, maintain, and deploy. This is because of the service provided by the top layer to the middle layer and partner of the relation, which should be avoided by the top layer. This would attract the attention of designer and developer, meaning the layers are not logically well separate. Similarly, the architecture has two skip-call from the top layer to the lower layer.

\begin{table}[h]
	\caption{Analysis of relation(arch) between the nodes of sample network 2.}
	\begin{center}
	\begin{tabular}{llllllll}
		
		\textbf{No} & \textbf{Arch} & \textbf{Layer} & \textbf{S(e)} & \textbf{D(e)} & $\mathcal{L}_s$ & $\mathcal{L}_d$    & \textbf{Remark} \\\hline
		1      & 1-2     & 1-1   & 1    & 2    & 3          & 3          & Normal \\
		2      & 4-1     & 2-1   & 4    & 1    & 2          & 3          & Back   \\
		3      & 2-4     & 1-2   & 2    & 4    & 3          & 2          & Normal \\
		4      & 2-5     & 1-2   & 2    & 5    & 3          & 2          & Normal \\
		5      & 2-10    & 1-3   & 2    & 10   & 3          & 1          & Skip   \\
		6      & 3-6     & 1-2   & 3    & 6    & 3          & 2          & Normal \\
		7      & 3-8     & 1-3   & 3    & 8    & 3          & 1          & Skip   \\
		8      & 5-3     & 2-1   & 5    & 3    & 2          & 3          & Back   \\
		9      & 4-7     & 2-3   & 4    & 7    & 2          & 1          & Normal \\
		10     & 4-8     & 2-3   & 4    & 8    & 2          & 1          & Normal \\
		11     & 5-4     & 2-2   & 5    & 4    & 2          & 2          & Normal \\
		12     & 5-9     & 2-3   & 5    & 9    & 2          & 1          & Normal \\
		13     & 6-5     & 2-2   & 6    & 5    & 2          & 2          & Normal \\
		14     & 6-9     & 2-3   & 6    & 9    & 2          & 1          & Normal \\
		15     & 6-10    & 2-3   & 6    & 10   & 2          & 1          & Normal \\\hline
	\end{tabular}
	\label{tab:archanalysis}
	\end{center}
\end{table}

\begin{table}[h]
	\caption{Back-call and skip-call violations of sample network. }
	\begin{center}
	\begin{tabular}{llllll}
		\textbf{Id} & \textbf{layer} &\textbf{ bcc} & \textbf{bcr} & \textbf{scc} &\textbf{ scr }\\\hline
		1  & 1     & 0   & 1   & 0  & 0   \\
		2  & 1     & 0   & 0   & 1  & 0   \\
		3  & 1     & 0   & 1   & 1  & 0   \\\hline
		4  & 2     & 1   & 0   & 0  & 0   \\
		5  & 2     & 1   & 0   & 0  & 0   \\
		6  & 2     & 0   & 0   & 0  & 0   \\\hline
		7  & 3     & 0   & 0   & 0  & 0   \\
		8  & 3     & 0   & 0   & 0  & 1   \\
		9  & 3     & 0   & 0   & 0  & 0   \\
		10 & 3     & 0   & 0   & 0  & 1   \\\hline
		Total & -  & 2   & 2   & 2  & 2 
	\end{tabular}
	\end{center}
	\begin{center}
	\begin{tabular}{lllll}
		
		\textbf{Layer} & \textbf{bcc} &\textbf{ bcr} & \textbf{scc} & \textbf{scr} \\\hline
		1  & 0   & 2   & 2  & 0   \\
		2  & 2   & 0   & 0  & 0   \\
		3  & 0   & 0   & 0  & 2  \\\hline
		Total & 2& 2   & 2  & 2
	\end{tabular}
	\label{tab:back-callandskip-call}
	\end{center}
\end{table}

\begin{table*}
	\caption{Node level analysis of sample network 2.}
	\begin{center}
	\begin{tabular}{|p{0.4in}|p{0.4in}|p{0.4in}|p{0.6in}|p{0.4in}|p{0.4in}|p{0.5in}|p{0.5in}|p{0.5in}|p{0.6in}|p{0.6in}| }
		\hline 
		\textbf{Node Id} & \textbf{In-Deg} & \textbf{Out-Deg} & \textbf{Degree} &\textbf{BVC}    & \textbf{BVR}   & \textbf{TBV}    & \textbf{SVC}    & \textbf{SVR}    &\textbf{ TSV}     & \textbf{AV}                                               \\\hline
		1    & 1   & 1    & 2   & -      & 1/1=1 & 1/2=.5 & 0/1=0  & -      & 0/2=0   & 1/2=.5                                            \\
		2    & 1   & 3    & 4   & -      & 0/1=0 & 0/4=0  & 1/3=.3 & -      & 1/4=.25 & 1/4=.25                                           \\
		3    & 1   & 2    & 3   & -      & 1/1=1 & 1/3=.3 & 1/2=.5 & -      & 1/3=.3  & 2/3=0.6                                           \\\hline
		4    & 2   & 3    & 5   & 1/3=.3 & 0/2=0 & 1/5=.2 & -      & -      & -       & 1/5=.2                                            \\
		5    & 2   & 3    & 5   & 1/3=.3 & 0/2=0 & 1/5=.2 & -      & -      & -       & 1/5=.2                                            \\
		6    & 1   & 3    & 4   & 0/3=0  & 0/1=0 & 0/4=0  & -      & -      & -       & 0/4=0                                             \\\hline
		7    & 1   & 0    & 1   & 0      & -     & 0      & -      & 0/1=0  & 0/1=0   & 0/1=0\\
		8    & 2   & 0    & 2   & 0      & -     & 0      & -      & 1/2=.5 & 1/2=.5  & 1/2=.5                                            \\
		9    & 2   & 0    & 2   & 0      & -     & 0      & -      & 0/2=0  & 0/2=0   & 0/2=0                                             \\
		10   & 2   & 0    & 2   & 0      & -     & 0      & -      & 1/2=.5 & 1/2=.5  & 1/2=.5  \\\hline                                         
	\end{tabular}
	\label{tab:nodelevel}
\end{center}
\end{table*}

\begin{table*}
	\caption{Layer wise analysis of sample network 2.}
	\begin{center}
	\begin{tabular}{|p{0.4in}|p{0.4in}|p{0.4in}|p{0.6in}|p{0.4in}|p{0.4in}|p{0.5in}|p{0.5in}|p{0.5in}|p{0.6in}|p{0.6in}| }
		\hline 
		\textbf{Layer ID} & \textbf{In-Deg} & \textbf{Out-Deg} & \textbf{Degree} & \textbf{BVC}    & \textbf{BVR}    & \textbf{TBV}     & \textbf{SVC}     & \textbf{SVR}     &\textbf{TSV}     & \textbf{AV} \\\hline
		1 & 3  & 6  & 9 &  -      & 2/3=.6 & 2/9=.2   & 2/6=.3  & -  	  & 2/9=0.2 & 4/9=.45 \\  
		2 & 5  & 9  & 14& 2/9=.2  & 0/5=0  & 2/14=.15 &  -      & -   	  & -       & 2/14=.15  \\
		3 & 7  & 0  & 7 & 0       & -      & 0/7=0    &  -      & 2/7=.29 & 2/7=.29 & 2/7=.29 \\ \hline 
		
	\end{tabular}
	\label{tab:layerlevel}
	\end{center}
\end{table*}
		
\par Three cycles are present in the sample network: 1) 1 - 2 - 4
2) 1 - 2 - 5 - 4  3) 3 - 6 - 5. The strongest component of sample network includes a set of nodes $N'= \{1,4,2,5,6,3\}$ with set of aches $|E'|= 8 $ and arches present in the strongest components are $\{1-2,2-5,5-3,3-6,6-5,5-4,4-1,2-4\}$. However, cyclic violation according to \cite{sarkar2009discovery} is 0.63 or 63\% because 5 out of 8 arches are crossing the layers. But only two back-call arches are critically involved in and completes the cycle while the rest of the arches are within the layer or follow the interaction rules. So cyclic violation according to our approach is only 0.25 or 25\% which much lower than measure in \cite{sarkar2009discovery} because our approach only considers the backward or skip calls in the cycle.

\begin{table}[h]
	\caption{Comparative analysis of both the approaches for sample network 2.}
	\begin{center}
	\begin{tabular}{cll}
		
		\textbf{Sr.} & \textbf{By approach} \cite{sarkar2009discovery}                                                                                          & \textbf{New approach}                                                           \\\hline
		1     & \begin{tabular}[c]{@{}l@{}}BACK =\{4,5\}\\ $L_{BACK}=\{2\}$\end{tabular}                                       & \begin{tabular}[c]{@{}l@{}}BV=\{4-1,5-3\}\\ BACK=S(BV)=\{4,5\}\\ \\ $L_{BACK}=\mathcal{L}(BV)$=\{2\}\end{tabular}          \\
		2     & \begin{tabular}[c]{@{}l@{}}BCVI(l):\\ l=1 \\ BCVI(1)=0/3=0\\ l=2\\ BCVI(2)=2/3=0.66\\ l=3\\ BCVI(3)=0/4=0\end{tabular} & \begin{tabular}[c]{@{}l@{}}BVL(l):\\l=1\\ BVL(1)=2/9=0.22\\ l=2\\ BVL(2)=2/14=0.15\\ l=3\\ BVL(3)=0/7=0\end{tabular} \\
		3     & \textbf{BCVI(S)=1-0.67=0.33}                                                                              & \textbf{BVS(S)=1-2/15=0.67}                                                                                            \\\hline
		4     & \begin{tabular}[c]{@{}l@{}}SKIP=\{2,3\}\\ $L_{SKIP}=\{1\}$\end{tabular}                                        & \begin{tabular}[c]{@{}l@{}}SV=\{2-10,3-8\}\\ SKIP=S(SV)=\{2,3\}\\ $L_{SKIP}=\mathcal{L}(SV)=\{1\}$\end{tabular}            \\
		5     & \begin{tabular}[c]{@{}l@{}}SCVI(L):\\ l=1\\ SCVI(1)=2/3=0.66\\\quad\\\quad\end{tabular}                                 & \begin{tabular}[c]{@{}l@{}}SVL(L):\\ l=1\\ SVL(1)=2/9=0.22\\l=3\\ SVL(3)=2/7=0.29\end{tabular}                  \\\hline
		6     & \textbf{SCVI(S)=1-2/3=0.33}                                                                                        & \textbf{SVS(S)=1-2/15=0.87}                                                                                            \\\hline
		7     & $DCVI(m^{msc})$=5/8=0.63                                                                                             & CV(msc)=2/8=0.25 \\\hline                                                                                            
	\end{tabular}
	\label{tab:comparative}
	\end{center}
\end{table}
\par The comparative of both the approaches have been shown in Table \ref{tab:comparative}. The average system violation for different setting of  $\alpha$ and $\beta$ is shown in Table \ref{tab:avgsystem}. However, first four entries in the table \ref{tab:avgsystem} represent an average system violation of strict or closed layering which in the range of 0.18 to 0.35. Further, last three entries in the Table \ref{tab:avgsystem} for open or soft layering allows skip-call violation, and are apparently found smaller value than closed layering.

\begin{table}[h]
	\caption{Average system violation for sample network 2.}
	\begin{center}
	\begin{tabular}{cllc}
		\textbf{Sr. No}	& \textbf{$\alpha$} & \textbf{$\beta$} & \textbf{AVS} \\ 
		\hline 
		1 & 0.5 & 0.5 & 0.18 \\ 
		2 & 0.75 & 0.5 & 0.24 \\ 
		3 & 1 & 0.5 & 0.29 \\ 
		4 & 1 & 1 & 0.35 \\\hline 
		5 & 0.5 & 0 & 0.12 \\ 
		6 & 0.75 & 0 & 0.18 \\ 
		7 & 1 & 0 & 0.24 \\ 
		\hline 
	\end{tabular} 
	\label{tab:avgsystem}
	\end{center}
\end{table}
\par The logical separation index for various layers is shown in Table \ref{tab:logiclasep}. Apparently we observer that lower layer is more logically separate than remaining two layers because no incoming and outgoing arches violate the back-call principles. It is also noticeable that top two layers have same numerator but different logical separation index due to different degree of the layers. 

\begin{table}[h]
	\caption{Logical separation index of layers of sample network 2.}
	\begin{center}
	\begin{tabular}{clllll}
		
		\textbf{Layer(l)} & \textbf{bcc} & \textbf{bcr} & \textbf{Deg} & \textbf{LS(l)} 		& \textbf{LSI(l)}  \\\hline
		1  		 & 0   & 2   & 9   & 2/9=0.22   & 1-0.22=0.78   \\
		2  		 & 2   & 0   & 14  & 2/14= 0.14 & 1-0.14=0.86   \\
		3 		 & 0   & 0   & 7   & 0/7=0  	& 1-0=1			\\\hline
	\end{tabular}
	\label{tab:logiclasep}
	\end{center}
\end{table}

\subsection{Experimental result of software systems}	
\textbf{ConStore}: ConStore is a small Java-based library used to model the real world problem using concept network. It facility to create and store the concepts. First, a meta model is needed to defined which describes the schema of various concepts. A model of real world problem is created based on the rules defined by the meta model. Model is expressed using graph in which nodes are concepts and edges represents relations between the concepts. Further, a facility to query the network to retrieve the part of network and allow to navigate the network is provided \cite{constore}. The comparative analysis of both approach is shown in Table \ref{tab:comparative_con_hw}.

\begin{figure}[h]
	\centering
	\includegraphics[width=1\linewidth, height=0.3\textheight]{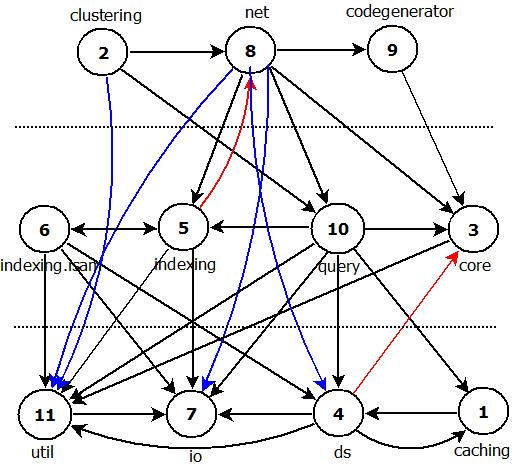}
	\caption{Package dependency network of Constore.}
	\label{fig:constorepack}
\end{figure}

\begin{figure*}
	\centering
	\includegraphics[width=1\linewidth, height=7cm]{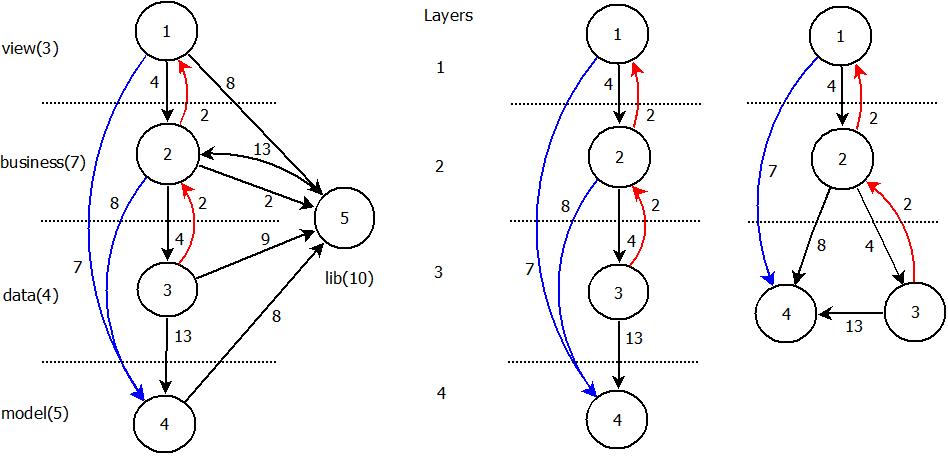}
	\caption{Layered structure of Health Watcher Application.}
	\label{fig:hwarch1}
\end{figure*}

\textbf{Health Watcher Application(HW)}:
A real web-based health management system implemented in Java, to improve the quality of the health services provided by health organizations. It has various facilities for citizens to post and update their health-related complaints, and so in response, they take required actions\cite{greenwood2007impact}. The architecture of HW is shown in Figure \ref{fig:hwarch1}, first illustrates the dependency amongst the packages, secondly represents the organization of packages into the four-layer structure and finally, the last figure is a three-layered structure which is most commonly referred structure in Java. The comparative analysis of both approaches is shown in the Table \ref{tab:comparative_con_hw}.

 \begin{table}
  \centering
  \caption{Comparative analysis of Constore}
  \begin{tabular}{|p{0.5in}|p{2in}p{2in}|}
   \hline 
   & \textbf{Approach defined in} \cite{sarkar2009discovery}                                                                                      & \textbf{New approach} \\ \hline 
 
 1                                 & \begin{tabular}[c]{@{}l@{}}BACK =\{4,5\}\\ $L_{BACK} = \{2,3\}$\end{tabular} & \begin{tabular}[c]{@{}l@{}}BV=\{5-8,4-3\}\\ BACK=S(BV)=\{4,5\}\\ $L_{BACK}= 
\mathcal{L}(BV)$=\{2,3\}\end{tabular}  \\ \hline 
	2                                 & \begin{tabular}[c]{@{}l@{}}BCVI(l): \\ l=2\\ BCVI(2)=1/4=0.25\\ l=3 \\ BCVI(3)=1/4=0.25\end{tabular}                     & \begin{tabular}[c]{@{}l@{}}BVL(l):\\ l=1 \\ BVL(1)=1/19=0.09\\ l=2 \\ BVL(2)=2/34=0.06\\ l=3 \\ BVL(3)=1/33=0.03\end{tabular} \\ \hline 
 
 	3                                 & \textbf{BCVI(S)=1-0.25=0.75}                                                                                             & \textbf{\begin{tabular}[c]{@{}l@{}}BVS(S)=1-2/43=0.096 			\end{tabular}}  \\ \hline 
 	
 		4                                 & \begin{tabular}[c]{@{}l@{}}SKIP=\{2,8\}\\ $L_{SKIP}=\{1\}$\end{tabular}                                                  & \begin{tabular}[c]{@{}l@{}}SV=\{8-7,2-11,8-11,8-4\} \\ SKIP=S(SV)=\{2,8\} \\ $L_{SKIP}=\mathcal{L}(SV)=\{1\}$\end{tabular} \\                                                                                                                                                   
		5                                 & \begin{tabular}[c]{@{}l@{}}SCVI(L):\\  l=1\\ SCVI(1)=2/3=0.66\end{tabular}                                               & \begin{tabular}[c]{@{}l@{}}SVL(L): \\ l=1\\ SVL(1)=4/19=0.21\\ l=3\\ SVL(3)=4/33=0.12\end{tabular}   \\
		6                                 & \textbf{SCVI(S)=1-2/3=0.33}                                                                                              & \textbf{\begin{tabular}[c]{@{}l@{}}SVS(S)=1-4/43=0.91			\end{tabular}}                                    \\ \hline
		7                                 & \begin{tabular}[c]{@{}l@{}}Component1				\\ $DCVI(m^{msc})$=6/12=0.5\\ Component2\\ $DCVI(m^{msc})$=4/8=0.5\end{tabular} & \begin{tabular}[c]{@{}l@{}}Component1			\\ $CV(msc_1)$=2/12=0.17			\\ Component2			\\ $CV(msc_2)$=2/8=0.25\end{tabular}                                                                     \\ \hline
		8                                 & -                                                                                                                        & \begin{tabular}[c]{@{}l@{}}LSI(l):			\\ l=1			\\ LS(1)=1/19=0.05\\ LSI(1)=1-0.05=0.95\\ l=2		\\ LS(2)=2/34=0.06\\ LSI(2)=1-0.06=0.94\\ l=3\\ LS(3)=1/33=0.03\\ LSI(3)=1-0.03=0.97\end{tabular}\\ \hline

\end{tabular}
 \end{table}

\begin{table*}
	\centering
  \caption{Comparative analysis of HW}
  \begin{tabular}{|p{0.5in}|p{2in}p{2in}|}
   \hline 
   & \textbf{Approach defined in} \cite{sarkar2009discovery}                                                                                      & \textbf{New approach} \\ \hline 
		1     & \begin{tabular}[c]{@{}l@{}}BACK =\{2,3\}\\ L\_\{BACK\} = \{1, 2\}\end{tabular}                      & \begin{tabular}[c]{@{}l@{}}BV=\{2-1,3-2\}\\ BACK=S(BV)=\{2,3\}\\ $L_{BACK}=\mathcal{L}(BV)$=\{1,2\}\end{tabular}     \\
		2                             & \begin{tabular}[c]{@{}l@{}}BCVI(l):\\  l=2\\ BCVI(2)=2/6=0.33\\ l=3\\ BCVI(3)=1/4=0.25\end{tabular} & \begin{tabular}[c]{@{}l@{}}BVL(l):\\ l=1\\ BVL(1)=2/19=0.10\\ l=2\\ BVL(2)=4/34=0.11\\ l=3\\ BVL(3)=2/57=0.03\end{tabular}                                                                              \\ 
		3                                     & \textbf{BCVI(S)=1-0.29=0.71}                                                                        & \textbf{\begin{tabular}[c]{@{}l@{}}BVS(S)=1-4/55=0.92
		\end{tabular}}  \\ \hline
		4                                & \begin{tabular}[c]{@{}l@{}}SKIP=\{1\}\\ $L_{SKIP}=\{1\}$\end{tabular}                               & \begin{tabular}[c]{@{}l@{}}SV=\{1-4\} \\ SKIP=S(SV)=\{1\} \\ $L_{SKIP}=\mathcal{L}(SV)=\{1\}$\end{tabular}                                                                                              \\
		5                                    & \begin{tabular}[c]{@{}l@{}}SCVI(L):\\  l=1\\ SCVI(1)=2/3=0.66\end{tabular}                          & \begin{tabular}[c]{@{}l@{}}SVL(L): \\ l=1\\ SVL(1)=7/19=0.37\\ l=3\\ SVL(3)=7/57=0.12\end{tabular}                                                                                                      \\
		6                                  & \textbf{SCVI(S)=1-2/3=0.33}                                                                         & \textbf{\begin{tabular}[c]{@{}l@{}}SVS(S)=1-7/55=0.87	\end{tabular}}                                                                                                                     \\ \hline
		7             & \begin{tabular}[c]{@{}l@{}}Component1				\\ $DCVI(m^{msc})$=5/7=0.71\end{tabular}                   & \begin{tabular}[c]{@{}l@{}}Component1				\\ $CV(msc_1)$=3/7=0.43\end{tabular}                                                                                                                           \\ \hline
		8     & -                                                                                                   & \begin{tabular}[c]{@{}l@{}}LSI(l):				\\ l=1				\\ LS(1)=2/19=0.10 \\ LSI(1)=1-0.10=0.90\\ l=2				\\ LS(2)=4/34=0.11\\ LSI(2)=1-0.11=0.89 \\ l=3			\\ LS(3)=2/57=0.03\\ LSI(3)=1-0.03=0.97\end{tabular}\\\hline
	\end{tabular}
	\label{tab:comparative_con_hw}
\end{table*}

\subsection{Practical Applications}
\par In this section, we demonstrate how these measure are applied to software systems at package level. 
To address issue 1 as mentioned above, we need to recover the layered structure of the software systems.
As we know that a layered architecture is prerequisite for detecting layering principle violations. For that reason either we must have artifact containing correct description of architecture of the software system  otherwise we have to extract it from the source of software system. In this experiment, due to unavailability of the design documents, we need to extract the architecture from source and then  determine violations.
\par For this reason, we have selected the layered recovery technique proposed by Belle et al.\cite{belle2014recovering} to show the practical applicability of proposed measures. The approach recovers the layered architecture of object oriented systems with help of clustering upon the responsibility tree built from package structure of software system. In particular, the authors first create the responsibility tree from naming information of the packages. Packages grouped to the same responsibility must have common starting portion in the namespace. This information is then used to create hierarchy of package structure and represented in the form tree termed as \textit{responsibility tree}. In second phase, the approach creates responsibility clustering by grouping the nodes in the tree with specified granularity of responsibilities.
\par
\begin{figure*}[!tbp]
  \begin{subfigure}[b]{0.5\textwidth}
    \includegraphics[width=\textwidth]{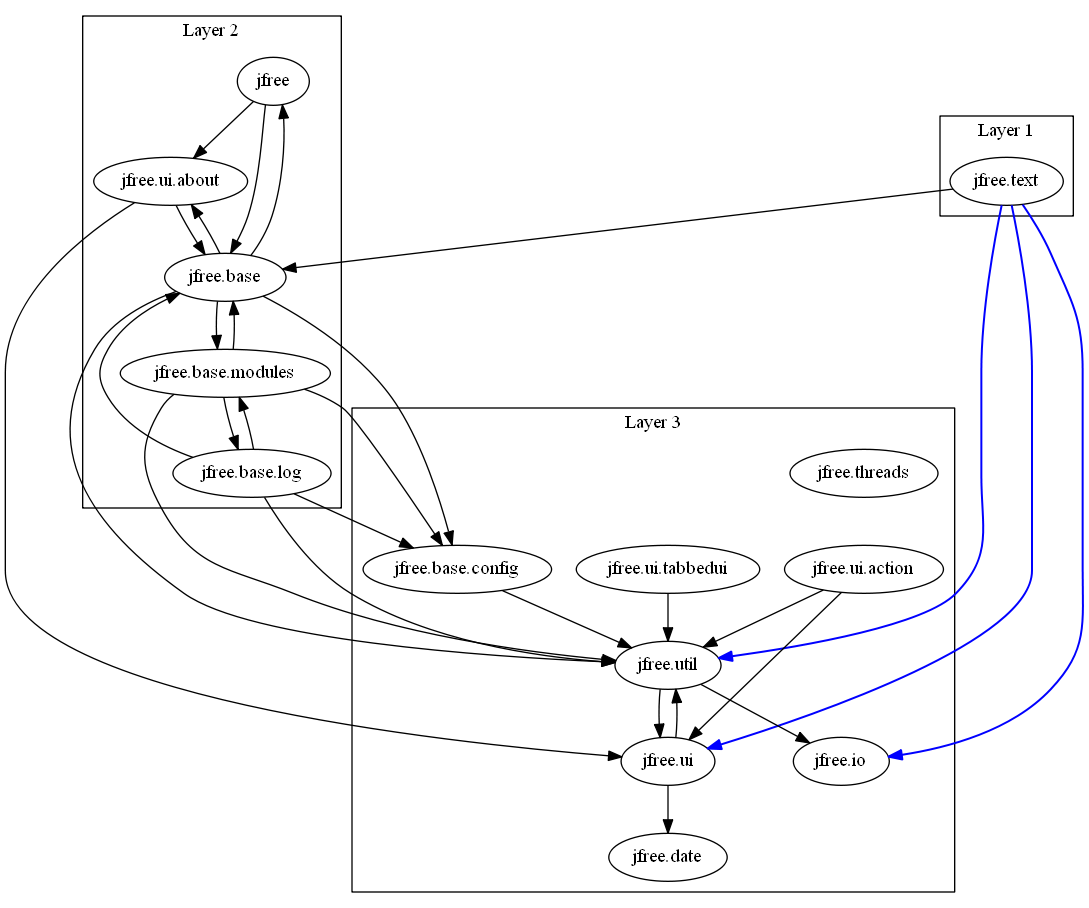}
    \caption{JCommon 1.0.0 library.}
    \label{fig:jcommon}
  \end{subfigure}
  \hfill
  \begin{subfigure}[b]{0.5\textwidth}
    \includegraphics[width=\textwidth]{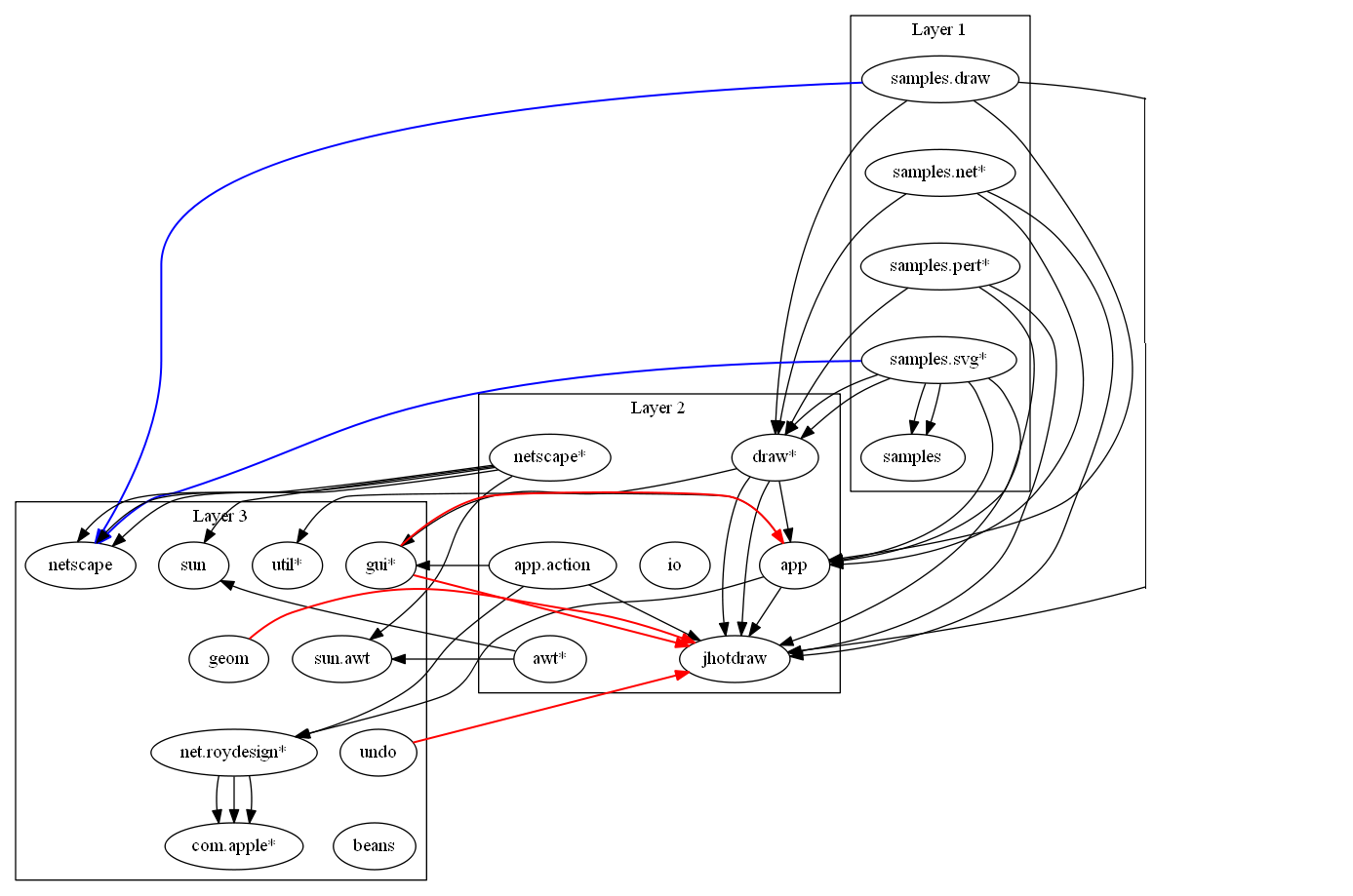}
    \caption{JHotDraw 7.0.6.}
    \label{fig:jhotdraw}
  \end{subfigure}
  \caption{Recovered layered architecture using Belle\cite{belle2014recovering} approach.}
\end{figure*}

 The approach applied to the three Java-based software system which are most commonly used in the reverse engineering. The three software system are JCommon, JHotDraw and JFreeChart. The layered architecture recovered from these software system using approach proposed by Belle\cite{belle2014recovering} is shown in Figure \ref{fig:jcommon}  and \ref{fig:jhotdraw}(Figure of JFreeChart is not included due to limitation of space).  

We analyzed the recovered layering architecture to address the issue 2 and 3.  The main aim of our assessment is to check the recovered architecture follows the layering style. If software has layered style then we further investigate two question 1) how good is recovered architecture(check violation of layering principles), and 2) what kind of layering style(open or closed) of recovered architecture.

\par We have applied the layered recovery approach to JCommon library, observed three level of responsibility and $8$ cluster with granularity $2$. In case with JCommon, the approach obtains best layering solution with only few violations as shown in Figure \ref{fig:jcommon}. In particular, three skip-call violations, and no back-call violations were detected in the recovered architecture.
At the same time, we notice the logical separation index for the architecture is $0.90$, which indicates that the recovered architecture is well separated. The cycles were present in the architecture but all are enclosed within layers, this means that no cycle span across the layer. Therefore, we conclude that the recovered architecture follows \textit{open layering} style.

\par In case of JHotDraw 7.0.6, we have observed the responsibility level is ranges to $5$ and system is well designed as it doesn't contain any cyclic dependency. We encounter few back-call violations in lower layer due the \textit{geom}, \textit{undo}, \textit{gui} packages call backward to \textit{app} and \textit{jhotdraw} packages. At same time, only two skip call violations that therefore the skip-call violation index is to be $0.096$. The logical separation index for JHotDraw is $0.93$ reveals a good layered structure and indicates that it follows \textit{open layering} style.

\par In case of JFreeChart 1.0.0, we have observed the responsibilities are grouped into 10 clusters and large quantity of  skip-call violations($51$) in the recovered architecture that calls to the lowest layer. We found no cycles in the recovered layered architecture.  The layers are well separated and therefore the logical separation index for all the layers is to be 1. Therefore, architecture follows a \textit{open layering} style because no back-violations is observed.

Here, we compare the results with previous approach.
Our approach provide better perspective of layer violations than the previous approach(Sarkar \cite{sarkar2009discovery}). It not only identify the package that violates the layering principles at same time it also detects dependency relations. These additional information is useful during simplification of the design and code for the developing team.
In case of JHotDraw, five bad relations contributed in back-call violations at same time surprisingly all indices are nearly equal. So, during simplification process we have to focus on resolving these relations.
In case of JCommon library and JFreeChart, we found relations which are permitted by the open layering style. But we need to do further investigations to verify the how close the recovered architecture to actual or intended design. However, the is not the scope of this paper but in future we will also explore this.

\par Hence, the results provided by our newly proposed measures are better than the existing measures owing to three main factors:
\begin{itemize}
    \item We do not just identify the violations but also obtain information about the relations that led to the same.
    \item This helps to gain better insights especially for code refactoring and simplification.
    \item It is more beneficial for the developer to know what relations to work upon rather than just locating what entities to work with.
\end{itemize}
\par Here, we address the issue of sensitivity of the proposed measures. The measures based on the concept of relationship are sensitive in nature because they are dependent on the accuracy of fact extracted from the source code. But the recently available tools for fact extractor like Dependency Finder(used in this projects) found to be accurately identified the dependency in source code, and we manually verified the fact extracted from the source code of software systems used in this experiment. At same time, all the measure are ratio of two quantities, where numerator represents bad relations and denominator is a total relations. It has been also observed that all bad relations are explicit relations in the form of inheritance, instantiation, and parameter, such relations are accurately captured by dependency extractor tool. So, even though the measure are sensitive, the accuracy are not compromised in any stage of the experiment.

\section{Conclusion}
\par In this paper, we have attempted to redefine the standard set of measures that are used to evaluate the layered architecture of the software. Along with this, we have defined measures to verify the logical separation of the layers to ensure the quality of the layered structure. We have observed that the relations significantly describe the various measure of layered architecture. The measures were used to evaluate five different software systems. The Constore application is well logically separated but the top two layers of HW are not well separated. At the same time, we have demonstrated how to use these measures in practical setting for three different software system. 
Future work includes the application of social network analysis or network theory to study and extract the layered architecture. These measures could then be applied for evaluation. As an extended scope, more large and complex systems can also be evaluated using these measures. The proposed measures are more succinct in nature and more inclusive of the domain-specific constraints, thus proving to be a better alternative as compared to the previous approaches.


\end{document}